\documentstyle[11pt,epsf]{article}
\newcommand{\be}{\begin{eqnarray}}
\newcommand{\ee}{\end{eqnarray}}

\title{\begin{flushright}
{\small SUNY-NTG-9637}
\end{flushright}
{\bf  QCD with Large Number of Quarks: \\Effects of
the  Instanton -- Anti-instanton Pairs.}}
\author{
 {\bf  M.Velkovsky and E.Shuryak } \\
{\it  Department of Physics,}\\
{\it State University of New York, Stony Brook, NY 11794}}

\begin{document}

\maketitle
\centerline{\bf Abstract}
We calculate the  contribution of the instanton -- anti-instanton ($I\bar I$)
pairs to the vacuum energy of QCD-like theories with $N_f$ light fermions 
using the saddle point method. We find a qualitative change of the behavior:
for $N_f \ge 6$ it starts to oscillate with  $N_f$.  Similar behaviour was
known for quantum mechanical systems interacting with fermions. We discuss the
possible consequences of this phenomenon, and its  relation to the mechanism of
chiral symmetry breaking in these theories. We also discuss the asymptotics of
the perturbative series associated with  the $I\bar I$ contribution, comparing
our results with those in literature.


\newpage                                                                  

In pure gauge theories the tunneling between topologically distinct classical
vacua (described semiclassically by instantons) are known to shift the ground
state energy $down$,  as in  quantum mechanics. It is often assumed that any
non-perturbative effects should do the same: e.g. the bag constant of the MIT
bag model was always assumed to be positive, without much discussion. However
this is no longer obvious if there are fermions in the theory. Moreover, in
supersymmetric theories  only $positive$ shifts, if any, are allowed. 

In order to see how transition from one regime to another may happen, Balitsky
and Yung \cite{Bal_Yung}  considered a toy model,  a quantum mechanical
double-well potential coupled to a fermion field with a  certain coupling 
constant $a$.  The contribution of  tunneling-antitunneling paths was found to
have a  complex phase proportional to  $a$,  which exactly flips sign at the
$a$ values   at which  the model becomes supersymmetric\footnote{The absolute
value of also coincides  with the known shift, which ensure that the method is
correct.}.

In this work we study what happens in the QCD-like gauge theories with $N_f$
light quarks. Unlike in the toy model mentioned, in this case  the instanton
contribution is only a part of the nonperturbative contribution to the vacuum
energy. However, there are multiple evidences  (see e.g. recent review
\cite{SS_96c}) that at least for $N_f=0-3$ (here and below we imply  that the
number of colors is $N_c=3$) it is a very important if not dominant part. The
so called ``instanton liquid'' model \cite{Shu_82}  picture the vacuum  of
those theories as a relatively dilute and uncorrelated      ensemble of
instantons: it reproduces many phenomenological facts about hadronic
correlations functions and spectroscopy, was recently directly supported by
multiple lattice studies, see e.g. \cite{lat}. With increasing temperature T 
correlations between instantons and anti-instantons become more important, so
that  the random picture is no longer valid.  Eventually a chiral restoration
phase transition takes place, in which only strongly correlated instanton -
anti-instanton pairs, or ``molecules'' \cite{IS2} are present. This idea
reproduces lattice data for close-to-QCD theories ($N_f=0-3$) on the position
and type of the chiral phase transitions, hadronic screening masses etc. 

The pattern and mechanism of the  chiral symmetry breaking at larger $N_f$ is
not yet understood. It is known that chiral symmetry should  be restored in the
vacuum above certain critical value $N_f^c$, which  should be  below the value
where the asymptotic freedom disappears $N_f^c<N_f^{AF}=11 N_c/2$.  Between
them the theory is conformal due to  the   infrared fixed point \cite{BZ}.
Studies of  the interacting instanton  ensemble \cite{SS_96} have shown that
instantons $alone$ cannot support the condensate already for $N_f>4$. Lattice
simulations  for $N_f=8$  \cite{Nf8}  has  found at weak coupling regime a
chirally symmetric phase,  similar to what  was recently observed for  $N_f=16$
\cite{Nf16}. Furthermore, for $N_f=4$ \cite{Columbia_4f} have the quark
condensate was found to be nonzero but drastically smaller than for $N_f=0-3$.
However,  estimates based on one loop gap equation \cite{ATW_96} suggest much
larger critical value $N_f^c\approx 11$: presumably the condensate induced by
this mechanism is too small be seen in the lattice measurements mentioned.

As $N_f$ is increased, the two basic components of the instanton ensemble
mentioned above, (i) ``single'' (uncorrelated) ones  and (ii) strongly
correlated $\bar I I $ pairs or ``molecules", have the opposite trends. Singles 
can only exist due to nonzero value of the quark condensate. The fermionic
determinant  proportional to $(|<\bar q q>| \rho^3)^{N_f}$) suppresses
small-size instantons, so that only  large ones, with sizes $\rho \sim |<\bar q
q>|^{-1/3}$, survive\footnote{ Those are  outside the semi-classical domain,
unless we are close to chiral restoration boundary \cite{AS_97}.}. 

The  ``molecules" do not need a nonzero quark condensate: thus they become the
dominant effect. Furthermore, as noticed in \cite{Shu_87b}, for increasing
$N_f$ they move  toward smaller sizes, and for $N_f>11N_c/2-3$ their density
even becomes $ultraviolet$ divergent\footnote{ It can be seen on purely
dimensional ground: the density of molecules is $dn_m\sim d\rho
\Lambda^{2b}\rho^{2b-5}$, which becomes UV divergent at $b<2$. This phenomenon
is similar to the UV divergence in the $O(3)$ $\sigma$ model: both are curious
examples of the UV divergences of $nonperturbative$ nature.}.  The differential
contribution to the normalized molecular partition  function of instantons and
anti-instantons with small (in order to use semi-classical formulae)  radii
$\rho_I,\rho_{\bar I}<<1/\Lambda$:
\be
{d^2 Z_{mol} \over d \rho_I d \rho_{\bar I}}=V^{(4)}C^2\rho^4\int d^4(R) 
d\Omega(-\rho^2 |T_{I\bar I}(R,\Omega)|^2)^{N_f}\exp(-S_{int}
(R,\Omega)),
\ee
where $ R $ is the separation in units of $\rho$ and $\Omega$ the relative 
orientation of the $I\bar I$ pair. $C$ is standard single 
instanton density:
\be
C= {4.6 \exp(-1.86 N_c) \over \pi^2 (N_c-1)!(N_c-2)!}{1 \over \rho^5}
(S_0)^{2N_c}\exp(-S_0),
\ee
Up to two loops the relation between the instanton action $S_0$ and size is 
\be
\rho=e^{-{S_0\over b}}|{2b\over b_1}S_0 +1|^{b1 \over 2b^2},
\ee
 where $b={11\over 3 }N_c-{2\over 3} N_f$, 
$b_1={34 \over 3} N_c^2-{13 \over 3}N_c N_f +{N_f \over N_c}$.
The so called overlap matrix element of the Dirac operator enteing here is
\be
T_{I \bar I}=\int dt d^3{\bf x}
\phi_I^\dagger(x-z_I,\Omega_I)
\not\!\!D\phi_{\bar I}(x-z_{\bar I},\Omega_{\bar I})
\ee
with $\phi$ being fermionic zero modes of the insatnton and anti-instanton.

  The gauge interaction $S_{int}(R,\Omega))$ in general depends on the
gauge configurations used. For
$\bar I I$ configurations one cannot use the 
equations of motion because there is no notrivial minimum:
 one should use  the Streamline equation instead
\cite{Bal_Yung,Yung,Ver_91}.
 However,  we
will only be interested in the saddle points at large
enough R, so that well known dipole formula would be sufficient.

The integrand in (1) has a maximum at $R \to 0$, so the integral appears to be
dominated by the weak field  configurations, which belong to perturbative
sector. The way to separate the non-perturbative physics is to use the saddle
point method,  moving the contour in the complex $R$ plane\cite{Bal_Yung}. Let
us split the contour into  2 parts, one going from $R=0$ to $R=i\infty$,
corresponding to the  perturbative contribution, and the other one going from
that point to $R=+\infty$ in such a way that $|R|$ is always  big, so that (4)
is applicable\footnote{Presumably there are no singularities that can prevent
it.}. The second integral is identified as a non-perturbative part of
$Z_{mol}$. It is generally complex, and its imaginary part should be
compensated by the perturbative contribution: thus the well known relation to
perturbative series:
\be
E_k^{pert}=-{1\over \pi} \int_0^\infty {dg\over g^{k+1}} Im(E^{non.pert}(g)),
\ee

At the saddle point  there is a  balance between the gauge and fermionic part
of the action:  therefore we treat $N_f$ as a largeparameter, which however
should also be  much smaller than the single instanton action $S_0$ in order to
ensure that the saddle point is at large R: $S_0>>N_f>>1$. If so, we may use
asymptotic  expressions for $T_{I\bar I}$ and $S_{int}$ \footnote{Note that the
addition of more terms in the expansion of $T_{I\bar I}$ and $S_{int}$ leads to
new saddle points at smaller separations. Those however are artefacts of the
truncation of the asymptotic series and should be disregarded. }. Changing to
variables $\beta=1/R^2$ and $\xi=\cos^2(\theta)$, where $\theta$ is the only
relevant orientation angle,  (4) becomes:
$$
{d^2 Z_{mol}^{non.pert} \over d \ln \rho_I d \ln \rho_{\bar
I}V^{(4)}}=C^2\rho^6 I 
$$
\be
I=\int_0^\pi {d\theta \sin^2(\theta) \over \pi/2 }  
4 \pi \int_0^\infty dR
R^3 e^{-S_04(1-4\cos^2\theta)}\left(-{16 \cos^2(\theta) \over R^6} \right)
^{N_f}
\cr = 4 (-16)^{N_f} \int_0^1 d\xi \sqrt{1-\xi \over \xi}
\xi^{N_f}\int_0^\infty {d\beta \over
\beta}\exp(-S_04\beta^2(1-4\xi)+(3N_f-2)\ln \beta).
\ee
There are two saddle points at 
$
\beta_0=\pm \sqrt{3N_f-2 \over 8S_0(1-4 \xi)}.
$
When $\xi < 1/4$ they are real and we take the positive one, but for 
$\xi > 1/4$ they are imaginary. It is irrelevant which one we 
choose (the choice is related to the definition of the perturbation series), 
so we take the one in the lower complex plane. Doing the $\beta$ integral by 
the steepest descent method (the answer is analytic in $\xi$), we get for 
(8):
\be
I=4 (-16)^{N_f}\sqrt{\pi}(8 S_0 e)^{-{3N_f-2\over 2}}(3N_f-2)^{3N_f-3\over 2}
\int_0^1 d\xi \sqrt{ 1-\xi \over \xi} \xi^{N_f} (1-4\xi)^{-{3N_f-2\over 2}}
\ee
The above integral has a singularity at $\xi=1/4$\footnote{ It is  actually an
artefact, due  to the truncation of $\beta= 1/R^2$ series. In fact close to
that point the  steepest descent method is not applicable if we truncate the
gauge action to order $O(\beta^3)$, because there is no large parameter in the
exponent. However the $O(\beta^4)$ term is nonzero at  $\xi=1/4$ and if we take
it into account, the integral will be finite. In this case, however, we have 4
saddle points, and we have the difficulty of defining the integral so that no
spurious ones contribute to it.} which can be avoided  by taking  $\xi$
integration over a contour in the complex plane. A contour in the upper $\xi$
plane matches the real positive  $\beta_0$ for $\xi<1/4$ to the negative
imaginary one we have chosen for $\xi>1/4$.

For even $N_f$ the integral (10) can be written as a contour integral around
the cut from 0 to 1 (avoiding $\xi=1/4$), and the contributions are from two
poles: $\xi=1/4$, which is of order $(3/2) N_f-1$ and gives purely imaginary
contribution and $\xi=\infty$, which is of order $3-1/2N_f$, which gives a real
contribution, but ceases to exist for $N_f \ge 6$. That is where the behaviour
of the integral changes. For odd $N_f$ one can express the integral via the
incomplete elliptic integrals and their derivatives. In this case there are
both real and imaginary contributions.  Table 1. contains  the values of the
real and imaginary parts of ${d^2 E_{mol.gas} \over d \ln\rho_I  d
\ln\rho_{\bar I}}$ in units $\Lambda$, for $N_c=3$, for $\rho={1 \over 3} {1
\over \Lambda} $ \footnote{ Note that for small $N_f$ the results are  
unreliable because  the condition $N_f >>1$ is not true; those are shown for
comparison  only. Note also that we have not attempted to integrate over the 
instanton sizes, because that cannot be done without introducing a particular
assumptions about the mechanism of the cutoff at large sizes.}.

\vskip .7 cm

{\bf Table 1.}
\vskip .3 cm

\vbox{\offinterlineskip
\halign{\strut \vrule \ \hfil # \hfil \ & \vrule \ \hfil # \hfil \ & \vrule 
\ \hfil # \hfil \ & \vrule \ \hfil # \hfil \ \vrule \cr
\noalign{\hrule}
$N_f$ & $Re {d^2 E_{mol.gas} \over d \ln\rho_I d \ln\rho_{\bar I}} 
 $ & $Im {d^2 E_{mol.gas} \over d \ln\rho_I d \ln\rho_{\bar I}} $ & $I(N_f) $
\cr
\noalign{\hrule}
1& $.7959*10^{-5}$&$-.1717*10^{-4}$&$-.7173*10^{-5}$ \cr  
\noalign{\hrule}                                                          
2&            0.  &$-.5098*10^{-6}$&$-.2690*10^{-5}$ \cr  
\noalign{\hrule}                                                          
3&$-.3668*10^{-7}$&$-.5474*10^{-7}$&$-.2833*10^{-5}$ \cr  
\noalign{\hrule}                                                          
4&$-.1917*10^{-7}$&$-.5330*10^{-8}$&$-.1991*10^{-5}$ \cr   
\noalign{\hrule}                                                          
5&$-.1027*10^{-7}$&$ .6464*10^{-8}$&$ .1178*10^{-4}$ \cr  
\noalign{\hrule}                                                          
6&            0.  &$ .1520*10^{-7}$&$ .7913*10^{-4}$ \cr
\noalign{\hrule}
7&$ .5067*10^{-7}$&$ .1127*10^{-7}$&$ .7147*10^{-4}$ \cr
\noalign{\hrule}
8&            0.  &$-.2714*10^{-5}$&$-.2292*10^{-2}$ \cr
\noalign{\hrule}
9&$-.1898*10^{-3}$&$ .1368*10^{-4}$&$ .2067*10^{-2}$ \cr
\noalign{\hrule}
10&            0. &$ .3365*10^{-4}$&$ .4120        $ \cr
\noalign{\hrule}
11&$.1093*10^{-6}$&$ .2421*10^{-8}$&$ .1480        $ \cr
\noalign{\hrule}
12&            0. &$-.6030*10^{-11}$&$ - 132.8     $ \cr
\noalign{\hrule}
}}

Here $I(N_f)= Im {d^2 E_{mol.gas} \over d \ln\rho_I d \ln\rho_{\bar I}} e^{2
S_0} S_0^{-4N_c+(3/2)N_f-1}$ is the imaginary part of the energy derivative 
with the dependence on $S_0$ (respectively $g$) removed. 

The obtained  $N_f$ dependence of the  instanton-- anti-instanton contribution
to the partition function is very peculiar. First of all, in many cases the
real part of the obtained  result  vanishes. Second, its  $signs$  start to
oscillate for $N_f>5$. Coincidentally, this is also the region where the 
instantons  cannot support chiral condensate, and, as discussed above, their
contribution is probably very small.

In this paper we have not investigated the effective interaction between
fermions  induced by the ``molecules'', but comment that  one should expect for
intermediate $N_f$  appearance of non-perturbative forces and hadronic states
having  {\it two distinct scales}, dictated by the condensate and
``molecules''. Furthermore (at least in the saddle point approximation used in
this work) their sign and magnitude of the latter should be proportional to the
vacuum shifts evaluated above. As a result, the oscillatory behaviour found in
this work should also propagate to correlators (at sufficient small distances)
and hadronic spectra.
 
Let us now evaluate  the large order  coefficients in the perturbation series
due instanton-- anti-instanton contribution. Using (7) we can find   the
energy:
\be
{d^2E_k^{pert}\over d \ln \rho_1 d \ln \rho_2}\rho^4=-{1\over \pi} 
\int {dg\over g^{k+1}} {8 \pi^2 \over 
g^2}^{4N_c-(3/2)Nf+1} e^{-{16 \pi^2 \over g^2}} I(N_f) \cr
=-I(N_f){1\over 2 \pi} \left({ 1\over 8 \pi^2}\right)^{k/2} \int_0^\infty 
d\left(S_0 \right) \left( 
S_0 \right)^{4 N_c-(3/2)N_f+k/2}e^{-2S_0},
\ee
This is yet another saddle point integral with a saddle point at 
$S_0=2N_c-3/4N_f+k/4$.
The applicability condition now is $k>>3N_f$. Finally we get the expected
factorial behaviour\footnote{
 The coefficients are smaller than those due to
``renormalons'', which however correspond to a completely different
set of diagrams, with maximal number of loops instead of classical
``trees''.}:
$$
{d^2E_k^{pert}\over d\ln \rho_1 d\ln \rho_2}\rho^4= -I(N_f)\sqrt{1\over 4 \pi }
\left({1 \over 8 \pi^2}\right)^{k/2} \times  
$$
\be
\left({4N_c-(3/2)N_f+k/2 \over 2}\right)
^{4 N_c-(3/2)N_f+k/2+1/2}e^{-(4N_c-(3/2)N_f+k/2)} \cr
=-{I(N_f)\over 2^{4N_c-(3/2)N_f}}\left({1 \over 4 \pi}\right)
^{k+1 } \Gamma(4N_c-(3/2)N_f+k/2+1).
\ee
One can compare it with the  results for the $\bar I I$ contribution to cross
section of the process of $e^+e^-\rightarrow hadrons$. In \cite{Sil} the 
factorial behavior is $(4N_c+k/2)!$ for  $N_f=N_c=3$. In \cite{SilFal} for the
standard choices of $N_c$ and $N_f$ the authors get
\be
R_{e^+e^-\rightarrow hadrons}=\sum -813(3280.5 k)^{-35/k}(10+k/2)! {g\over 4
\pi}^k , 
\ee
which is quite close to the behaviour of (11) for $N_c=3$ and $N_f=2$.

In summary, we  have calculated the contribution of a correlated instanton--
anti-instanton pair to the partition function of the QCD-like theories with
increasing number of fermions. On general grounds (and the analogy to the
quantum-mechanical problem with fermions) it was expected that the behaviour
should have an oscillatory pattern, shifting the ground state down or upward as
$N_f$ grows. This was indeed found to be true, but only for large enough $N_f$.
The same is expected to happen for the hadronic spectra. Our results for the
asymptotic of the perturbative series  generated by $\bar I I$ configurations
are very close to those obtained in \cite{Sil,SilFal}  by a different method.

%\bibliographystyle{plain}  
%\bibliography{rev}  

\end{document}